\documentclass[11pt]{article}
\usepackage[utf8]{inputenc}
\usepackage{mathpazo}
\usepackage{leftindex}
\usepackage{bm}
\allowdisplaybreaks
\usepackage{aurical}
\usepackage[toc,page]{appendix}
\usepackage{latexsym,amsfonts,amsmath,amssymb,mathrsfs,bbold,mathtools,esint,amsthm,mathtools}
\usepackage{dsfont}
\usepackage{mathtools}
\usepackage{float}
\usepackage{graphicx}
\usepackage{hyperref}
\usepackage{setspace}
\usepackage{color}
\usepackage{slashed}
\usepackage{cleveref}
\numberwithin{equation}{section}
\usepackage[affil-it]{authblk}

\usepackage{float}
\usepackage{indentfirst}
\usepackage{soul}
\usepackage{booktabs}
\usepackage{eso-pic,graphicx}
\usepackage{nicefrac}
\usepackage{epsfig}

\newcommand{\str}{\mathrm{Str}}

\newcommand{\mi}{\mathrm{i}\,}
\usepackage{braket}
\def\tsc#1{\csdef{#1}{\textsc{\lowercase{#1}}\xspace}}

\usepackage[bottom=0.5cm, right=1.5cm, left=1.5cm, top=1.5cm]{geometry}

\usepackage{color,hyperref}
\definecolor{darkblue}{rgb}{0.0,0.0,0.3}
\hypersetup{colorlinks,breaklinks,linkcolor=darkblue,urlcolor=darkblue,anchorcolor=darkblue,citecolor=darkblue}

\usepackage{graphicx, nicefrac}
\usepackage[numbers,sort&compress]{natbib}
\newcommand{\jae}{\mathrm{e}}
\newcommand{\jasn}{\mathrm{sn}}
\newcommand{\jacn}{\mathrm{cn}}
\newcommand{\ccos}{\mathsf{c}}
\newcommand{\ssin}{\mathsf{s}}

\DeclareMathOperator{\csch}{csch}
\usepackage{braket}
\def\tsc#1{\csdef{#1}{\textsc{\lowercase{#1}}\xspace}}
\tsc{WGM}
\tsc{QE}
\tsc{EP}
\tsc{PMS}
\tsc{BEC}
\tsc{DE}
\usepackage{palatino}

\title{On integrability of the one-dimensional Hubbard model} 

\author[$\dagger$]{A. Melikyan\footnote{\href{mailto:amelik@gmail.com}{amelik@gmail.com}}}
\affil[$\dagger$]{Centro Internacional de Física\\ Instituto de Física\\
Universidade de Brasília\\
70910-900, Brasília, DF, Brasil}

\begin{document}
\maketitle
\let\WriteBookmarks\relax
\def\floatpagepagefraction{1}
\def\textpagefraction{.001}

\begin{abstract}
We find a family of solutions to Zamolodchikov's tetrahedral algebra corresponding to the fermionic $R$-operator for the free fermion model of the difference type in one of the spectral parameters,  construct an extension of the $R$-operator for a system of two spins satisfying the Yang-Baxter equation, and find the local charges. We also construct a twisted monodromy operator, which leads to the one-dimensional Hubbard model.
\end{abstract}
\section{Introduction}
We consider the fermionic $R$ operator, which corresponds to the three-parameter elliptic parameterization of the $R$-matrix of the free-fermion model, introduced in \cite{Bazhanov1984, Bazhanov:1984ji, Bazhanov:1984jg} (see also \cite{Baxter:86ch,Baxter:1987eq} for the relation between the free fermion model and the checkerboard Ising model\footnote{I thank Prof. Jacques H.H. Perk for drawing my attention to this result.}), and can be written (after applying the Jordan-Wigner transformation) in the following form \cite{Melikyan:2022nen}:
\begin{align}
    R_{ab}(u_{a}-u_{b},&\zeta_{a},\zeta_{b};k)=a_{0}
(u_{ab},\zeta_{a},\zeta_{b};k) + a_{1}
(u_{ab},\zeta_{a},\zeta_{b};k)\mathsf{n_{a}} +a_{2}
(u_{ab},\zeta_{a},\zeta_{b};k) \mathsf{n_{b}}+a_{3}(u_{ab},\zeta_{a},\zeta_{b};k)\mathsf{n_{a}}\mathsf{n_{b}} \notag \\ 
&+ c_{1}(u_{ab},\zeta_{a},\zeta_{b};k)\mathsf{\Delta_{ab}}+c_{2}(u_{ab},\zeta_{a},\zeta_{b};k)\mathsf{\Delta_{ba}}+d_{1}(u_{ab},\zeta_{a},\zeta_{b};k)\mathsf{\tilde{\Delta}^{(+)}_{ab}}+d_{2}(u_{ab},\zeta_{a},\zeta_{b};k)\mathsf{\tilde{\Delta}_{ab}} \label{intro:Rop_orig},
\end{align}
where we have introduced the shorthand notation $u_{ab}=u_{a}-u_{b}$, the relation between the functions in \eqref{intro:Rop_orig} and the coefficients of the $R$-matrix \cite{Bazhanov1984} (written in terms of the Jacobi elliptic functions with the elliptic modulus $k$) is given in Appendix \ref{app:nots}, and we have defined the following operators:
\begin{align}
&\mathsf{n_{a}}=\mathsf{c^{\dagger}_{a}c_{a}},\, \mathsf{\Delta_{ab}}=\mathsf{c^{\dagger}_{a}c_{b}},\,
\mathsf{\tilde{\Delta}^{(+)}_{ab}}=\mathsf{c^{\dagger}_{a}c^{\dagger}_{b}},\,
\mathsf{\tilde{\Delta}_{ab}}=\mathsf{c_{a}c_{b}},
\notag \\
&\{\mathsf{c^{\dagger}_{a}},\mathsf{c_{b}}\}_{+}=\delta_{ab},\,
\{\mathsf{c_{a}},\mathsf{c_{b}}\}_{+}=0,
\{\mathsf{c^{\dagger}_{a}},\mathsf{c^{\dagger}_{b}}\}_{+}=0.
\label{intro:nots}
\end{align}
The $R$-operator \eqref{intro:Rop_orig} satisfies the Yang-Baxter equation:
\begin{align}
    R_{ab}(u_{ab},\zeta_{a},\zeta_{b};k)R_{ac}(u_{ac},\zeta_{a},\zeta_{c};k)R_{bc}(u_{bc},\zeta_{b},\zeta_{c};k)=R_{bc}(u_{bc},\zeta_{b},\zeta_{c};k)R_{ac}(u_{ac},\zeta_{a},\zeta_{c};k)R_{ab}(u_{ab},\zeta_{a},\zeta_{b};k),\label{intro:YBE_orig}
\end{align}
together with the decorated Yang-Baxter equation \cite{Melikyan:2022nen}:
\begin{align}
   R_{ab}(u_{ab},\zeta_{a},\zeta_{b}-2K(k);-k)(2\mathsf{n_{a}}-1) R_{ac}(u_{ab},\zeta_{a},\zeta_{c}-2K(k);-k)R_{bc}(u_{bc},\zeta_{b},\zeta_{c};k)=\notag \\
   R_{bc}(u_{bc},\zeta_{b},\zeta_{c};k)R_{ac}(u_{ab},\zeta_{a},\zeta_{c}-2K(k);-k)(2\mathsf{n_{a}}-1) R_{ab}(u_{ab},\zeta_{a},\zeta_{b}-2K(k);-k),\label{intro:decoYBE}
\end{align}
where $K(k)$ is the complete elliptic integral of the first kind. Using \eqref{intro:YBE_orig} and \eqref{intro:decoYBE} we had found in \cite{Melikyan:2022nen}, in the trigonometric limit, the most general solution for the arbitrary values of the spectral parameters $\{u_{a},\zeta_{a}\}$ to Zamolodchikov's tetrahedral algebra \cite{Zamolodchikov1981,Korepanov1993,Korepanov2013,Korepanov1994b,Korepanov:1989tl,Baxter1986321,Bazhanov2012}. This means solving the following equations for the coefficients $W^{\alpha \beta \gamma}_{\lambda \mu \nu }$:\footnote{The lengthy list of all equations arising from \eqref{intro:Zam_tetral}, as well as the explicit form of the corresponding solution for the coefficients $W^{\alpha \beta \gamma}_{\lambda \mu \nu }$ is given in the appendices in \cite{Melikyan:2022nen}. We also stress that the coefficients $W^{\alpha \beta \gamma}_{\lambda \mu \nu }$ in Zamolodchikov's algebra span, in general, a six-dimensional space (cf. \cite{Umeno1998b}).}
\begin{align}
\mathcal{L}^{\alpha}_{ab}\mathcal{L}^{\beta}_{ac}\mathcal{L}^{\gamma}_{bc}=\sum_{\lambda,\mu,\nu=0}^{1} W^{\alpha \beta \gamma}_{\lambda \mu \nu }\mathcal{L}^{\lambda}_{bc}\mathcal{L}^{\mu}_{ac}\mathcal{L}^{\nu}_{ab},\label{intro:Zam_tetral}
\end{align}
where $\{\alpha, \beta, \gamma\}=\{0,1\}$, and the operators $\mathcal{L}^{\alpha}$ are defined as follows:
\begin{align}
    \mathcal{L}^{0}_{ab}(u_{ab},\zeta_{a},\zeta_{b};k)&=R_{ab}(u_{ab},\zeta_{a},\zeta_{b};k),\label{intro:L0}\\
     \mathcal{L}^{1}_{ab}(u_{ab},\zeta_{a},\zeta_{b};k)&=R_{ab}(u_{ab},\zeta_{a},\zeta_{b}-2K(k);-k) (2\mathsf{n_{a}}-1)\label{intro:L1}
\end{align}

Our main interest to study this particular parameterization of the $R$ operator \eqref{intro:Rop_orig} arises from the fact that it is of the difference type in one of the spectral parameters, namely, with respect to the $\{u_{a}\}$ parameters, while the dependence on the other parameters ($\{\zeta_{a}\}$) is arbitrary. Thus, the $R$ operator \eqref{intro:Rop_orig} is a suitable candidate for understanding the quantization of fermionic relativistic non-ultralocal integrable systems, such as the massive fermion model appearing in string theory on a $AdS_{5} \times S^{5}$ background \cite{Alday:2005jm}. It is easy to show that fixing the $\{\zeta_{a}\}$ parameters, and considering the Yang-Baxter equation \eqref{intro:YBE_orig} with respect to $\{u_{a}\}$, one obtains the $XY$ model in an external field, controlled by the $\{\zeta_{a}\}$ parameters.\footnote{In addition, the Lax connection was also obtained in \cite{Melikyan:2020ibw}, allowing the investigation of non-ultralocality in the continuous limit.} On the other hand, the $S$ matrix of string theory on the $AdS_{5} \times S^{5}$ background has been shown to be related to the $R$-matrix of the one-dimensional Hubbard model \cite{Essler:2005bk,Beisert:2006qh,Rej2006,Mitev2017}, which is not Lorentz invariant \cite{Frahm:1990ab, Links:2001ab}. In this paper we show that the $R$-operator \eqref{intro:Rop_orig} is also suitable for obtaining the one-dimensional Hubbard model. To this end, in Section \ref{ext} we construct an extension of the $R$ operator for a system with two spins and use the general solution to Zamolodchikov's tetrahedral algebra found in \cite{Melikyan:2022nen} to show that it satisfies the Yang-Baxter equation. Then, in Section \ref{hub}, considering the Yang-Baxter equation \eqref{intro:YBE_orig} with respect to the parameters $\{\zeta_{a}\}$ (while fixing the parameters $\{u_{a}\}$), we find the local charges and construct a twisted monodromy operator, which leads to the one-dimensional Hubbard model.

\section{Extended $R$-operator}
\label{ext}
An immediate consequence of the general solution to Zamolodchikov's algebra \eqref{intro:Zam_tetral}\footnote{ In this and the next sections we consider the trigonometric limit $k \to 0$.}, as was shown in \cite{Melikyan:2022nen}, is the possibility to extend the fermionic $R$-operator to the following form:
\begin{align}
    R^{\textrm{(ext)}}_{ab}(u_{ab},\zeta_{a},\zeta_{b};k)=\mathcal{L}^{0}_{ab}(u_{ab},\zeta_{a},\zeta_{b};k) +\left(c_{ab}\right)\mathcal{L}^{1}_{ab}(u_{ab},\zeta_{a},\zeta_{b};k).\label{ext:single_spin_extension}
\end{align}
The coefficients $c_{ab}=c_{ab}(u_{ab},\zeta_{a},\zeta_{b})$ were determined to have the form: $c_{ab}=\tanh\left[h_{a}(u_{a},\zeta_{a})-h_{b}(u_{b},\zeta_{b})\right]$ (where $h_{a}(u_{a},\zeta_{a})$ are arbitrary functions) from the 
requirement that the operator $R^{\textrm{(ext)}}$ satisfies the Yang-Baxter equation.

We now extend this result to systems with two spins $\sigma=\ket{\uparrow},\, \ket{\downarrow}$. To this end, we look for an extension of the form (cf. \cite{Essler:2005bk,Umeno1998,Umeno1998b,Umeno2000,Shiroishi1995as,Shiroishi1995-ae}):
\begin{align}
    R^{\textrm{(ext)}}_{ab}(u_{ab},\zeta_{a},\zeta_{b};k)=\left(b_{ab}\right)\mathcal{L}^{0 ({\uparrow})}_{ab}(u_{ab},\zeta_{a},\zeta_{b};k) \mathcal{L}^{0 ({\downarrow})}_{ab}(u_{ab},\zeta_{a},\zeta_{b};k)+\left(c_{ab}\right)\mathcal{L}^{1 ({\uparrow})}_{ab}(u_{ab},\zeta_{a},\zeta_{b};k)\mathcal{L}^{1 ({\downarrow})}_{ab}(u_{ab},\zeta_{a},\zeta_{b};k).\label{ext:two_spin_extension}
\end{align}
where $b_{ab}=b_{ab}(u_{ab},\zeta_{a},\zeta_{b})$ and $c_{ab}=c_{ab}(u_{ab},\zeta_{a},\zeta_{b})$ are some functions to be determined, as above, from the condition that $R^{\textrm{(ext)}}$ satisfies the Yang-Baxter equation:
\begin{align}
     R^{\textrm{(ext)}}_{ab}(u_{ab},\zeta_{a},\zeta_{b}) R^{\textrm{(ext)}}_{ac}(u_{ac},\zeta_{a},\zeta_{c}) R^{\textrm{(ext)}}_{bc}(u_{bc},\zeta_{b},\zeta_{c})= R^{\textrm{(ext)}}_{bc}(u_{bc},\zeta_{b},\zeta_{c}) R^{\textrm{(ext)}}_{ac}(u_{ac},\zeta_{a},\zeta_{c}) R^{\textrm{(ext)}}_{ab}(u_{ab},\zeta_{a},\zeta_{b}),\label{ext:YBE_two_spin}
\end{align}
In the most general case, this results in 36 non-linear equations for $(b_{ab},c_{ab})$. Unlike the single-spin case \eqref{ext:single_spin_extension}, the general analysis of \eqref{ext:two_spin_extension} is much more involved, given the complexity of the $W^{\alpha \beta \gamma}_{\lambda \mu \nu }$ functions in Zamolodchikov's algebra \eqref{intro:Zam_tetral}, and we postpone it to future work. In what follows, we will consider a particular case, which is, however, enough for our purposes. Namely, we will look for a set of spectral parameters which reduce the general six-dimensional space, spanned by $W^{\alpha \beta \gamma}_{\lambda \mu \nu }$, to a lower-dimensional space.\footnote{I thank \cite{Melikyan:2022nen}'s anonymous referee for raising the problem of comparing solutions to Zamolodchikov algebra found in \cite{Melikyan:2022nen} to those found, e.g., in \cite{Umeno1998b}. We recall that the latter spans a three-dimensional space.}

We start with the following case: we fix the parameters $u_{a}=u_{b}=u_{c} \equiv u$ and let the parameters $(\zeta_{a},\zeta_{b},\zeta_{c})$ be arbitrary. Using the appendices of \cite{Melikyan:2022nen} this leads to the following solution to \eqref{intro:Zam_tetral} (we list only non-vanishing coefficients):
\begin{equation}\label{ext:W_equal_us}
\begin{aligned}[c]
W^{010}_{001}&=\frac{\sin\left[\frac{\zeta_{a}-\zeta_{b}}{2}\right] \sin\left[\zeta_{c}\right]}{\sin\left[\frac{\zeta_{a}-\zeta_{c}}{2}\right] \sin\left[\frac{\zeta_{b}+\zeta_{c}}{2}\right]};\\
W^{100}_{001}&=\frac{\sin\left[\frac{\zeta_{b}-\zeta_{c}}{2}\right] \sin\left[\frac{\zeta_{a}+\zeta_{c}}{2}\right]}{\sin\left[\frac{\zeta_{a}-\zeta_{c}}{2}\right] \sin\left[\frac{\zeta_{b}+\zeta_{c}}{2}\right]};\\
W^{001}_{001}&=-\frac{\sin\left[\frac{\zeta_{a}-\zeta_{b}}{2}\right] \sin\left[\zeta_{c}\right]}{\sin\left[\frac{\zeta_{a}-\zeta_{c}}{2}\right] \sin\left[\frac{\zeta_{b}+\zeta_{c}}{2}\right]};\\
 W^{111}_{001}&=-\frac{\cos\left[\frac{\zeta_{b}-\zeta_{c}}{2}\right] \cos\left[\frac{\zeta_{a}+\zeta_{c}}{2}\right]}{\sin\left[\frac{\zeta_{a}-\zeta_{c}}{2}\right] \sin\left[\frac{\zeta_{b}+\zeta_{c}}{2}\right]};
\end{aligned}
\quad 
\begin{aligned}[c]
W^{010}_{010}&=-\frac{\sin\left[\frac{\zeta_{a}-\zeta_{b}}{2}\right] \sin\left[\frac{\zeta_{b}-\zeta_{c}}{2}\right]}{\sin\left[\frac{\zeta_{a}+\zeta_{b}}{2}\right] \sin\left[\frac{\zeta_{b}+\zeta_{c}}{2}\right]};\\
W^{100}_{010}&=\frac{\sin\left[\frac{\zeta_{a}+\zeta_{c}}{2}\right] \sin\left[\zeta_{b}\right]}{\sin\left[\frac{\zeta_{a}+\zeta_{b}}{2}\right] \sin\left[\frac{\zeta_{b}+\zeta_{c}}{2}\right]};\\
W^{001}_{010}&=\frac{\sin\left[\frac{\zeta_{a}+\zeta_{c}}{2}\right] \sin\left[\zeta_{b}\right]}{\sin\left[\frac{\zeta_{a}+\zeta_{b}}{2}\right] \sin\left[\frac{\zeta_{b}+\zeta_{c}}{2}\right]};\\
W^{111}_{010}&=-\frac{\cos\left[\frac{\zeta_{a}-\zeta_{b}}{2}\right] \cos\left[\frac{\zeta_{b}-\zeta_{c}}{2}\right]}{\sin\left[\frac{\zeta_{a}+\zeta_{b}}{2}\right] \sin\left[\frac{\zeta_{b}+\zeta_{c}}{2}\right]};
\end{aligned}
\quad
\begin{aligned}[c]
W^{010}_{100}&=\frac{\sin\left[\frac{\zeta_{b}-\zeta_{c}}{2}\right] \sin\left[\zeta_{a}\right]}{\sin\left[\frac{\zeta_{a}+\zeta_{b}}{2}\right] \sin\left[\frac{\zeta_{a}-\zeta_{c}}{2}\right]};\\
W^{100}_{100}&=-\frac{\sin\left[\frac{\zeta_{b}-\zeta_{c}}{2}\right] \sin\left[\zeta_{a}\right]}{\sin\left[\frac{\zeta_{a}+\zeta_{b}}{2}\right] \sin\left[\frac{\zeta_{a}-\zeta_{c}}{2}\right]};\\
 W^{001}_{100}&=\frac{\sin\left[\frac{\zeta_{a}-\zeta_{b}}{2}\right] \sin\left[\frac{\zeta_{a}+\zeta_{c}}{2}\right]}{\sin\left[\frac{\zeta_{a}+\zeta_{b}}{2}\right] \sin\left[\frac{\zeta_{a}-\zeta_{c}}{2}\right]};\\
 W^{111}_{100}&=\frac{\cos\left[\frac{\zeta_{a}-\zeta_{b}}{2}\right] \cos\left[\frac{\zeta_{a}+\zeta_{c}}{2}\right]}{\sin\left[\frac{\zeta_{a}+\zeta_{b}}{2}\right] \sin\left[\frac{\zeta_{a}-\zeta_{c}}{2}\right]}.
\end{aligned}
\end{equation}
In addition, we also obtain the following relation:
\begin{align}
    \mathcal{L}^{1}_{ab} \mathcal{L}^{1}_{ac} \mathcal{L}^{1}_{bc} = \left(Y^{111}_{001}\right)\mathcal{L}^{0}_{ab} \mathcal{L}^{0}_{ac} \mathcal{L}^{1}_{bc}+\left(Y^{111}_{010}\right)\mathcal{L}^{0}_{ab} \mathcal{L}^{0}_{ac} \mathcal{L}^{1}_{bc}+\left(Y^{111}_{100}\right)\mathcal{L}^{0}_{ab} \mathcal{L}^{0}_{ac} \mathcal{L}^{1}_{bc},\label{ext:L111_indep}
\end{align}
where:
\begin{align}
    Y^{111}_{001}=-\frac{\cot\left[\frac{\zeta_{a}-\zeta_{c}}{2}\right]}{\tan\left[\frac{\zeta_{a}+\zeta_{c}}{2}\right]};\quad
     Y^{111}_{001}=-\frac{\cot\left[\frac{\zeta_{a}+\zeta_{b}}{2}\right]}{\tan\left[\frac{\zeta_{b}+\zeta_{c}}{2}\right]};\quad
      Y^{111}_{001}=-\frac{\cot\left[\frac{\zeta_{a}-\zeta_{c}}{2}\right]}{\tan\left[\frac{\zeta_{a}+\zeta_{b}}{2}\right]}.\label{ext:Y_equal_us}
\end{align}
Substituting \eqref{ext:two_spin_extension} into \eqref{ext:YBE_two_spin}, and using the formulas \eqref{ext:W_equal_us}-\eqref{ext:Y_equal_us}, we find six independent nonlinear equations for $\{c_{ab}\}$, listed in Appendix \ref{app:eqs}.\footnote{Without loss of generality, we set here $b_{ab}=b_{ac}=b_{bc}=1$.} It can be shown that for the equations (\ref{app:eqs:eq1}-\ref{app:eqs:eq6}) to hold the following conditions must be satisfied:
\begin{align}
&\left\{c_{ab}\right\}\frac{\cos\left[\zeta_{a}\right]+\cos\left[\zeta_{b}\right]}{\sin\left[\zeta_{a}\right] \sin\left[\zeta_{b}\right]}+\left\{-c_{ac}\right\}\frac{\cos\left[\zeta_{a}\right]+\cos\left[\zeta_{c}\right]}{\sin\left[\zeta_{a}\right] \sin\left[\zeta_{c}\right]}+\left\{c_{bc}\right\}\frac{\cos\left[\zeta_{b}\right]+\cos\left[\zeta_{c}\right]}{\sin\left[\zeta_{b}\right] \sin\left[\zeta_{c}\right]}=0,\label{ext:maineq1}\\ \newline\notag \\
&\frac{\left(-1+\left(c_{ac}\right)^{2}\right)\cos\left[\zeta_{a}\right] +\left(1+\left(c_{ac}\right)^{2}\right)\cos\left[\zeta_{c}\right]}{c_{ac}\sin\left[\zeta_{c}\right]}-\frac{\left(-1+\left(c_{ab}\right)^{2}\right)\cos\left[\zeta_{a}\right]+\left(1+\left(c_{ab}\right)^{2}\right)\cos\left[\zeta_{b}\right]}{c_{ab}\sin\left[\zeta_{b}\right]}=0.\label{ext:maineq2}
\end{align}
We can solve these equations noting that due to the functional dependencies of the coefficients: $c_{ab}=c_{ab}(\zeta_{a},\zeta_{b})$, $c_{ac}=c_{ac}(\zeta_{a},\zeta_{c})$, $c_{bc}=c_{bc}(\zeta_{b},\zeta_{c})$, for the equation \eqref{ext:maineq2} itself to hold, the following compatibility conditions should be satisfied:
\begin{align}       
\frac{\left(-1+\left(c_{ac}\right)^{2}\right)\cos\left[\zeta_{a}\right] +\left(1+\left(c_{ac}\right)^{2}\right)\cos\left[\zeta_{c}\right]}{c_{ac}\sin\left[\zeta_{c}\right]}=f(\zeta_{a}),\label{ext:compat_eq1}\\
\frac{\left(-1+\left(c_{ab}\right)^{2}\right)\cos\left[\zeta_{a}\right]+\left(1+\left(c_{ab}\right)^{2}\right)\cos\left[\zeta_{b}\right]}{c_{ab}\sin\left[\zeta_{b}\right]}=f(\zeta_{a}),\label{ext:compat_eq2}
\end{align}
where the function $f(\zeta_{a})$ depends only on the parameter $\zeta_{a}$. To determine its precise form, we proceed as follows. First, solving equations \eqref{ext:compat_eq1} and \eqref{ext:compat_eq2} for $c_{ac}$ and $c_{ab}$, we find $c_{bc}$ from equation \eqref{ext:maineq1}. Then, the functional requirement $c_{bc}=c_{bc}(\zeta_{b},\zeta_{c})$ leads to the following differential equation for $f(\zeta_{a})$:
\begin{align}
f(\zeta_{a})\frac{d f(\zeta_{a})}{d \zeta_{a}}-\cot\left(\zeta_{a}\right)\left(4+f(\zeta_{a})\right)=0.\label{ext:dif_eq_f}
\end{align}
The general solution of \eqref{ext:dif_eq_f} has the form:
\begin{align}
    f(\zeta_{a})=2 \mi \sqrt{1-\lambda^{2} \sin^{2}\left(\zeta_{a}\right)},\label{ext:f_solution}
\end{align}
where $\lambda$ is an arbitrary complex constant. This fixes the functions $c_{ab}=c_{ab}(\zeta_{a},\zeta_{b})$, and the explicit form is easily found from the equations (\ref{ext:maineq1}-\ref{ext:compat_eq2}):\footnote{We note, that there is, in fact, a family of solutions, corresponding to different choices for the signs of the square roots, which will not be considered here.}
\begin{align}
    c_{ab}(\zeta_{a},\zeta_{b})=\mi \frac{-\sin\left[\zeta_{b} \right]\sqrt{1-\lambda^{2} \sin^{2}\left(\zeta_{a}\right)}+\sin\left[\zeta_{a} \right]\sqrt{1-\lambda^{2} \sin^{2}\left(\zeta_{b}\right)}}{\cos\left[\zeta_{a} \right]+\cos\left[\zeta_{b}\right]}.\label{ext:sol_cab_gen}
\end{align}
The functions for $c_{ac}=c_{ac}(\zeta_{a},\zeta_{c})$ and $c_{bc}=c_{bc}(\zeta_{b},\zeta_{c})$ have the same forms as in \eqref{ext:sol_cab_gen}. The simplest choice of the constant $\lambda=0$ then leads to the following solution (cf. \eqref{ext:single_spin_extension}):
\begin{align}
c_{ab}(\zeta_{a},\zeta_{b})=\mi \tan\left[\frac{\zeta_{a}-\zeta_{b}}{2} \right];\quad c_{ac}(\zeta_{a},\zeta_{c})=\mi \tan\left[\frac{\zeta_{a}-\zeta_{c}}{2} \right];\quad c_{bc}(\zeta_{b},\zeta_{c})=\mi \tan\left[\frac{\zeta_{b}-\zeta_{c}}{2} \right].\label{ext:lamda_zero}
\end{align}

In summary, we have found a solution for the extended $R$ operator \eqref{ext:two_spin_extension} that satisfies the Yang-Baxter equation \eqref{ext:YBE_two_spin}, by fixing the parameters $u_{a}=u_{b}=u_{c} \equiv u$ and allowing the parameters $(\zeta_{a},\zeta_{b},\zeta_{c})$ to be arbitrary. There are other solutions, and for the next section we will need the solution for the following case: $u_{a}=u_{b}=\pi+ u_{c}$. The analysis follows the same steps as described above, and here we give only the final formulas. For this case, it is convenient to take: $c_{ac}=c_{bc}=b_{ab}=1$ in \eqref{ext:two_spin_extension}. Then we find the following solution:
\begin{align}
    \hat{c}_{ab}(\zeta_{a},\zeta_{b})&=\mi \frac{-\sin\left[\zeta_{b} \right]\sqrt{1-\lambda^{2} \sin^{2}\left(\zeta_{a}\right)}+\sin\left[\zeta_{a} \right]\sqrt{1-\lambda^{2} \sin^{2}\left(\zeta_{b}\right)}}{\cos\left[\zeta_{a} \right]+\cos\left[\zeta_{b}\right]},\label{ext:sol_cab_gen_pi} \\ \newline \notag \\
    \hat{b}_{ac}(\zeta_{a},\zeta_{c})&=\mi \frac{\cos\left[\zeta_{a} \right]-\cos\left[\zeta_{c}\right]}{\sin\left[\zeta_{c} \right]\sqrt{1-\lambda^{2} \sin^{2}\left(\zeta_{a}\right)}+\sin\left[\zeta_{a} \right]\sqrt{1-\lambda^{2} \sin^{2}\left(\zeta_{c}\right)}},\label{ext:sol_bac_gen_pi} \\ \newline \notag \\
    \hat{b}_{bc}(\zeta_{b},\zeta_{c})&=\mi \frac{\cos\left[\zeta_{b} \right]-\cos\left[\zeta_{c}\right]}{\sin\left[\zeta_{c} \right]\sqrt{1-\lambda^{2} \sin^{2}\left(\zeta_{b}\right)}+\sin\left[\zeta_{b} \right]\sqrt{1-\lambda^{2} \sin^{2}\left(\zeta_{c}\right)}},\label{ext:sol_bbc_gen_pi} 
\end{align}
The choice $\lambda=0$ reduces the above solution to:
\begin{align}
\hat{c}_{ab}(\zeta_{a},\zeta_{b})=\mi \tan\left[\frac{\zeta_{a}-\zeta_{b}}{2} \right];\quad \hat{b}_{ac}(\zeta_{a},\zeta_{c})=\mi \tan\left[\frac{\zeta_{a}-\zeta_{c}}{2} \right];\quad \hat{b}_{bc}(\zeta_{b},\zeta_{c})=\mi \tan\left[\frac{\zeta_{b}-\zeta_{c}}{2} \right].\label{ext:lamda_zero_pi}
\end{align}
We note (and use this observation in the next section) that the expressions \eqref{ext:sol_cab_gen} for $c_{ab}(\zeta_{a},\zeta_{b})$ and \eqref{ext:sol_cab_gen_pi} for $\hat{c}_{ab}(\zeta_{a},\zeta_{b})$ coincide for both solutions.

\section{The local charges and the one-dimensional Hubbard model}
\label{hub}
We now calculate the local charges of the model and show that the solutions obtained in the previous section can be used to obtain the one-dimensional Hubbard model. First, we consider the solution \eqref{ext:sol_cab_gen} (i.e., we take $u_{a}=u_{b}=u_{c}$ and suppress everywhere below the dependence on $\{u_{a}\}$), and, thus, we have the following extended $R$-operator, satisfying the Yang-Baxter equation \eqref{ext:YBE_two_spin}:
\begin{align}
    R^{\textrm{(ext)}}_{ab}(\zeta_{a},\zeta_{b})=\mathcal{L}^{0 ({\uparrow})}_{ab}(\zeta_{a},\zeta_{b}) \mathcal{L}^{0 ({\downarrow})}_{ab}(\zeta_{a},\zeta_{b})+c_{ab}\left( \zeta_{a}, \zeta_{b}\right)\mathcal{L}^{1 ({\uparrow})}_{ab}(\zeta_{a},\zeta_{b})\mathcal{L}^{1 ({\downarrow})}_{ab}(\zeta_{a},\zeta_{b}),\label{hub:Rext_sol1}
\end{align}
where we have denoted $\mathcal{L}^{0,1 \, ({\sigma})}_{ab}(\zeta_{a},\zeta_{b})=\mathcal{L}^{0,1 \,({\sigma})}_{ab}(0;\zeta_{a},\zeta_{b})$.
We note, using \eqref{hub:Rext_sol1} and \eqref{ext:sol_cab_gen}, that $R^{\textrm{(ext)}}_{ab}(\zeta_{a},\zeta_{b})$ satisfies the following conditions:
\begin{align}
  R^{\textrm{(ext)}}_{ab}(\zeta_{a},\zeta_{b})  \big{\vert}_{\zeta_{a} \to\zeta_{b}}&=\left(\eta^{2}_{0}\right)\mathcal{P}^{(\uparrow)}_{ab}\mathcal{P}^{(\downarrow)}_{ab},\label{hub:initial_condition} \\
  \frac{d R^{\textrm{(ext)}}_{ab}(\zeta_{a},\zeta_{b})}{d \zeta_{a}} \big{\vert}_{\zeta_{a} \to\zeta_{b}}&=\left(\eta_{0}\right)\frac{d \mathcal{L}^{0 ({\uparrow})}_{ab}(\zeta_{a},\zeta_{b})}{d \zeta_{a}} \big{\vert}_{\zeta_{a} \to\zeta_{b}} \mathcal{P}^{(\downarrow)}_{ab} + \left(\eta_{0}\right) \mathcal{P}^{(\uparrow)}_{ab}\frac{d \mathcal{L}^{0 ({\downarrow})}_{ab}(\zeta_{a},\zeta_{b})}{d \zeta_{a}} \big{\vert}_{\zeta_{a} \to\zeta_{b}} \notag \\
  &+\frac{\mi}{2 \sqrt{1-\lambda^{2}\sin^{2}\left(\zeta_{b}\right)}}\mathcal{L}^{1 ({\uparrow})}_{ab}(\zeta_{b},\zeta_{b})\mathcal{L}^{1 ({\downarrow})}_{ab}(\zeta_{b},\zeta_{b}).\label{hub:der_Rext}
\end{align}
Here, $\mathcal{P}^{(\sigma)}$ is the fermionic permutation operator:
\begin{align}
    \mathcal{P}^{(\sigma)}_{ab}=1-{\mathsf{n}_{\mathsf{a},(\sigma)}} -{\mathsf{n}_{\mathsf{b},(\sigma)}}+{\mathsf{\Delta}_{\mathsf{ab},(\sigma)}}+{\mathsf{\Delta}_{\mathsf{ba},(\sigma)}},\label{hub:perm_op}
\end{align}
and $\eta_{0}$ is a constant (below we will set $\eta_{0}=1$ by an appropriate choice of the normalization $\rho$ in \eqref{app:nots:Rg_ij}).
Following the standard calculation (see, for example, \cite{Essler:2005bk,Umeno1998b} for more details), one defines the commuting quantities:
\begin{align}
   & \tau(\zeta_{a};\zeta_{1},\ldots,\zeta_{N})=\str_{a}\left[ R^{\textrm{(ext)}}_{aN}(\zeta_{a},\zeta_{N})R^{\textrm{(ext)}}_{a,N-1}(\zeta_{a},\zeta_{N-1}) \ldots R^{\textrm{(ext)}}_{a1}(\zeta_{a},\zeta_{1})\right],\label{hub:tau_def}\\
   &\left[\tau(\zeta_{a};\zeta_{1},\ldots,\zeta_{N}),  \tau(\zeta_{b};\zeta_{1},\ldots,\zeta_{N})\right]=0. 
\end{align}
Taking $\zeta_{N}=\zeta_{N-1}=\ldots=\zeta_{1} \equiv \zeta_{0}$, using \eqref{hub:initial_condition}, \eqref{hub:der_Rext} and Appendix \ref{app:nots}, we find the following Hamiltonian:\footnote{It is convenient to set here the normalization $\rho=\frac{\mi}{2\jae(\zeta_{0})\jasn(\zeta_{0})}$ in \eqref{app:nots:Rg_ij}. In this case, the constant $\eta_{0}=1$ in \eqref{hub:initial_condition}.}
\begin{align}
  \mathcal{H} &= \left(\eta_{1}\right) \, \, \tau^{-1}\left(\zeta_{0},\{\zeta_{0}\}\right) \left( \frac{d \tau}{d \zeta_{a}} \right)\big{\vert}_{\zeta_{a} \to \zeta_{0} = \mi z_{0}}\label{hub:charge} \\
  &=\left(\eta_{1}\right)\frac{\mi \csch(z_{0})}{2} \sum_{\substack{\sigma= \uparrow \downarrow}} \sum_{\substack{j=1}}^{N} \left[{\mathsf{\Delta}_{\mathsf{j+1,j} \, \sigma}}-{\mathsf{\Delta}_{\mathsf{j,j+1} \, \sigma}} \right]\notag\\
  &+\left(\eta_{1}\right) \frac{\mi \csch^{2}(z_{0})}{2 \sqrt{1+\lambda^{2}\sinh^{2}\left(z_{0}\right)}} \sum_{j=1}^{N} \left \{\left[ \mathsf{\Delta_{j,j+1\, \uparrow}} + \mathsf{\Delta_{j+1,j\, \uparrow}} \right]+\cosh(z_{0})\left[-1+n_{j}-n_{j+1}\right] + \sinh(z_{0})\left[-n_{j}+n_{j+1}\right]\right\} \notag \\ 
  &\qquad \qquad \qquad \qquad \qquad \qquad \times 
  \left \{\left[ \mathsf{\Delta_{j,j+1\, \downarrow}} + \mathsf{\Delta_{j+1,j\, \downarrow}} \right]+\cosh(z_{0})\left[-1+n_{j}-n_{j+1}\right] + \sinh(z_{0})\left[-n_{j}+n_{j+1}\right]\right\} \notag
\end{align}
where $\eta_{1}$ is an arbitrary constant that will be fixed later. Higher-order local charges can be found similarly by taking $n$th-order derivatives of $\ln(\tau)$ with respect to $\zeta_{a}$. The structure of the Hamiltonian \eqref{hub:charge} is similar to that of the generalized one-dimensional Hubbard model \cite{Essler:2005bk}, the essential difference being the opposite relative signs between $\Delta_{j,j+1 \, \sigma}$ and $\Delta_{j+1,j \, \sigma}$ in the first and second sums in \eqref{hub:charge}. In fact, we make the following observation: under the transformation (which leaves the operator algebra, as well as the vacuum invariant):\footnote{Here we ignore the boundary conditions.}
\begin{align}
    c_{j \, \sigma} \longrightarrow (\mi)^{j+1}c_{{j \, \sigma}},\label{hub:transform} 
\end{align}
the Hamiltonian \eqref{hub:charge} indeed corresponds to the generalized one-dimensional Hubbard model \cite{Essler:2005bk}.

To make this observation more precise, we now show that the one-dimensional Hubbard model can be obtained by considering a different family of local charges. To this end, we define the following twisted monodromy and transfer operators an:\footnote{We consider, for simplicity, the case  when the number of sites $N$ is an even number.}
\begin{align}
    &\widetilde{\mathcal{T}}_{a}(\zeta_{a};\{\zeta_{}\})=\widehat{R}^{\textrm{(ext)}}_{aN}(\zeta_{a},\zeta_{N})\mathcal{O}_{a}
    {R}^{\textrm{(ext)}}_{a,N-1}(\zeta_{a},\zeta_{N-1})\widehat{R}^{\textrm{(ext)}}_{a,N-2}\mathcal{O}_{a}(\zeta_{a},\zeta_{N-2}) \ldots \mathcal{O}_{a}{R}^{\textrm{(ext)}}_{a1}(\zeta_{a},\zeta_{1}),\label{hub:twisted_monop}\\
    &\tilde{\tau}(\zeta_{a};\{\zeta_{j}\})=\str_{a}\left[\widetilde{\mathcal{T}}_{a}(\zeta_{a};\{\zeta_{}\})\right],\label{hub:twisted_transop}
\end{align}
where for $j=1,3,\ldots,N-1$ we use the solution \eqref{ext:sol_cab_gen} (which corresponds to the case $u_{a}=u_{b}=u_{j}$) with ${R}^{\textrm{(ext)}}$ defined by \eqref{hub:Rext_sol1}, and for $j=2,4,\ldots,N$ we use the solution \eqref{ext:sol_cab_gen_pi}-\eqref{ext:sol_bbc_gen_pi} (which corresponds to the case $u_{a}=u_{b}=\pi + u_{j}$), where  $\widehat{R}^{\textrm{(ext)}}$ is defined by:
\begin{align}
     \widehat{R}^{\textrm{(ext)}}_{aj}(\zeta_{a},\zeta_{j})=\widehat{b}_{aj}\left( \zeta_{a}, \zeta_{j}\right)\widehat{\mathcal{L}}^{0 ({\uparrow})}_{aj}(\zeta_{a},\zeta_{j})\widehat{\mathcal{L}}^{0 ({\downarrow})}_{aj}(\zeta_{a},\zeta_{j})+\widehat{\mathcal{L}}^{1 ({\uparrow})}_{aj}(\zeta_{a},\zeta_{j}) \widehat{\mathcal{L}}^{1 ({\downarrow})}_{aj}(\zeta_{a},\zeta_{j}),\label{hub:Rext_sol2}
\end{align}
where $\widehat{\mathcal{L}}^{0,1 \, ({\sigma})}_{ab}(\zeta_{a},\zeta_{b})=\mathcal{L}^{0,1 \,({\sigma})}_{ab}(\pi;\zeta_{a},\zeta_{b})$. In addition,  we have inserted, after each even site, the operator $\mathcal{O}_{a}=(2n_{a \uparrow} -1)(2n_{a \downarrow} -1)$. As noted in the previous section, the expressions \eqref{ext:sol_cab_gen} for $c_{ab}(\zeta_{a},\zeta_{b})$ and \eqref{ext:sol_cab_gen_pi} for $\hat{c}_{ab}(\zeta_{a},\zeta_{b})$ coincide, and, as a result, also using the identity $\left[{R}^{\textrm{(ext)}}_{ab} ,\mathcal{O}_{a}\mathcal{O}_{b}\right]=0$, we obtain the commuting quantities: $\left[\tilde{\tau}(\zeta_{a};\{\zeta_{j}\},  \tilde{\tau}(\zeta_{b};\{\zeta_{j}\}\right]=0$. The formulas, analogous to \eqref{hub:initial_condition} and \eqref{hub:der_Rext} have the form:
\begin{align}
  \widehat{R}^{\textrm{(ext)}}_{aj}(\zeta_{a},\zeta_{j})  \big{\vert}_{\zeta_{a} \to\zeta_{j}}&=\left(\eta^{2}_{0}\right)\widehat{\mathcal{P}}^{(\uparrow)}_{aj}\widehat{\mathcal{P}}^{(\downarrow)}_{aj},\label{hub:initial_condition_hat} \\
 \frac{d \widehat{R}^{\textrm{(ext)}}_{aj}(\zeta_{a},\zeta_{j})}{d \zeta_{a}} \big{\vert}_{\zeta_{a} \to\zeta_{j}}&=\left(\eta_{0}\right)\frac{d \widehat{\mathcal{L}}^{1 ({\uparrow})}_{aj}(\zeta_{a},\zeta_{j})}{d \zeta_{a}} \big{\vert}_{\zeta_{a} \to\zeta_{j}} \widehat{\mathcal{P}}^{(\downarrow)}_{aj} + \left(\eta_{0}\right) \widehat{\mathcal{P}}^{(\uparrow)}_{aj}\frac{d \hat{\mathcal{L}}^{1 ({\downarrow})}_{aj}(\zeta_{a},\zeta_{j})}{d \zeta_{a}} \big{\vert}_{\zeta_{a} \to\zeta_{j}} \notag \\
  &+\frac{\mi}{2 \sqrt{1-\lambda^{2}\sin^{2}\left(\zeta_{j}\right)}}\mathcal{L}^{0 ({\uparrow})}_{aj}(\zeta_{a},\zeta_{j})\mathcal{L}^{0 ({\downarrow})}_{aj}(\zeta_{a},\zeta_{j}),\label{hub:der_hatRext}
\end{align}
where we have defined (cf. \eqref{hub:perm_op}) the following operators:
\begin{align}
\widetilde{\mathcal{{P}}}^{(\sigma)}_{aj} = 1-{\mathsf{n}_{\mathsf{a},(\sigma)}} -{\mathsf{n}_{\mathsf{j},(\sigma)}}+\mi {\mathsf{\Delta}_{\mathsf{aj},({\sigma})}}-\mi {\mathsf{\Delta}_{\mathsf{ja},(\sigma)}},\label{hub:perm_op_tilde} \\
   \widehat{\mathcal{{P}}}^{(\sigma)}_{aj} = 1-{\mathsf{n}_{\mathsf{a},(\sigma)}} -{\mathsf{n}_{\mathsf{j},(\sigma)}}-\mi {\mathsf{\Delta}_{\mathsf{aj},(\sigma)}}+\mi {\mathsf{\Delta}_{\mathsf{ja},(\sigma)}}.\label{hub:perm_op_hat} 
\end{align}
In addition, one can show that:
\begin{align}
&\tilde{\tau}(\zeta_{0};\{\zeta_{0}\})=\prod_{\substack{\sigma=\uparrow \downarrow}}\left(\widehat{\mathcal{{P}}}^{(\sigma)}_{12}\,\widetilde{\mathcal{{P}}}^{(\sigma)}_{23}\,\widehat{\mathcal{{P}}}^{(\sigma)}_{34}\ldots \,\widehat{\mathcal{{P}}}^{(\sigma)}_{N-1,N}\right);\quad \left(\widehat{\mathcal{{P}}}^{(\sigma)}_{ij}\right)^{2}=1;\quad \left(\widetilde{\mathcal{{P}}}^{(\sigma)}_{ij}\right)^{2}=1, \label{hub:twisted_tau0} \\ 
&\tilde{\tau}^{-1}(\zeta_{0};\{\zeta_{0}\})=\prod_{\substack{\sigma=\uparrow \downarrow}}\left(\widehat{\mathcal{{P}}}^{(\sigma)}_{N-1,N}\,\widetilde{\mathcal{{P}}}^{(\sigma)}_{N-2,N-1}\,\widehat{\mathcal{{P}}}^{(\sigma)}_{N-3,N-2}\ldots \,\widehat{\mathcal{{P}}}^{(\sigma)}_{12}\right).\label{hub:twisted_tau0inverse}
\end{align}
Using the above expressions, one can calculate the local charge:
\begin{align}
\widetilde{\mathcal{H}} &=\left(\eta_{1} \right)\tilde{\tau}^{-1}\left(\zeta_{0},\zeta_{0}\right) \left( \frac{d \tilde{\tau}}{d \zeta_{a}} \right)\big{\vert}_{\zeta_{a} \to \zeta_{0} = \mi z_{0}} \label{hub:final} \\
  &=\left(\eta_{1}\right)\frac{ \csch(z_{0})}{2} \sum_{\substack{\sigma= \uparrow \downarrow}} \sum_{\substack{j=1}}^{N} \left[{\mathsf{\Delta}_{\mathsf{j+1,j} \, \sigma}}+{\mathsf{\Delta}_{\mathsf{j,j+1} \, \sigma}} \right]\notag\\
  &+\left(\eta_{1}\right) \frac{\csch^{2}(z_{0})}{2 \sqrt{1+\lambda^{2}\sinh^{2}\left(z_{0}\right)}} \sum_{j=1}^{N} \left \{\left[ \mathsf{\Delta_{j,j+1\, \uparrow}} - \mathsf{\Delta_{j+1,j\, \uparrow}} \right]+\cosh(z_{0})\left[-1+n_{j}-n_{j+1}\right] + \sinh(z_{0})\left[-n_{j}+n_{j+1}\right]\right\} \notag \\ 
  &\qquad \qquad \qquad \qquad \qquad \qquad \times 
  \left \{\left[ \mathsf{\Delta_{j,j+1\, \downarrow}} - \mathsf{\Delta_{j+1,j\, \downarrow}} \right]+\cosh(z_{0})\left[-1+n_{j}-n_{j+1}\right] + \sinh(z_{0})\left[-n_{j}+n_{j+1}\right]\right\}, \notag
\end{align}
which corresponds to the generalized one-dimensional Hubbard model. To consider a specific example, we set: $\eta_{1}=\left[\frac{- \csch(z_{0})}{2}\right]^{-1}$, and calculate $\mathcal{H}_{0} = \lim\limits_{z_{0} \to \infty}\mathcal{H}$. The result is the one-dimensional Hubbard model:
\begin{align}
    \mathcal{H}_{0}=&-\sum_{\substack{\sigma= \uparrow \downarrow}} \sum_{\substack{j=1}}^{N} \left[{\mathsf{\Delta}_{\mathsf{j,j+1} \, \sigma}}+{\mathsf{\Delta}_{\mathsf{j+1,j} \, \sigma}}\right] +\left(\frac{1}{\lambda_{0}} \right)\sum_{j=1}^{N}\left \{2\mathsf{n_{j\, \uparrow}} - 1\right\}\left \{2\mathsf{n_{j\, \downarrow}} - 1\right\},\label{hub:Hub}
\end{align}
where $\lambda_{0}=\mi\lambda$.

To summarize, we have found: (i) a non-trivial solution to Zamolodchikov's tetrahedral algebra \eqref{intro:Zam_tetral} that results in the extended $R$ operator for a system of two spins, leading to the Hamiltonian \eqref{hub:charge} for the family of local charges corresponding to the transfer operator \eqref{hub:tau_def}, and (ii) the one-dimensional Hubbard model \eqref{hub:final}, which follows from the twisted monodromy operator \eqref{hub:twisted_monop}. It would be interesting to find more general solutions which allow nontrivial extensions for systems with two spins. Another interesting problem is to explore the relation of the free-fermion model with the checkerboard Ising model and its generalizations \cite{Baxter:86ch,Baxter:1987eq} to construct such extensions.

\section*{Acknowledgments}
I would like to thank Aleksandr Pinzul for the comments and useful discussions.

\appendix
\renewcommand{\thesection}{\Alph{section}.\arabic{section}}
\setcounter{section}{0}

\appendix
\section*{Appendices}
\section{$R$-matrix parametrization and gauge transformation}
\label{app:nots}
\setcounter{equation}{0}
\renewcommand\theequation{A.\arabic{equation}}
Here we give the formulas for the functions in the $R$-operator \eqref{intro:Rop_orig}. The coefficients of the original $R_{BS}$, given in \cite{Bazhanov1984, Bazhanov:1984ji, Bazhanov:1984jg} (also used in \cite{Melikyan:2022nen}), have branch cuts. In this paper, we use a more convenient gauge-transformed $ R^{(g)}$-matrix, the coefficients of which are meromorphic functions of the spectral parameters \cite{Bazhanov1984}. The gauge transformation has the form:
\begin{align}
    R^{(g)}(u_{ab},\zeta_{a},\zeta_{b};k)=g R_{BS} g^{-1},\label{app:nots:Gauge_transf}
\end{align}
where:
\begin{align}
g=\left[
    M\left(\sqrt{\jae(\zeta_{a}) \jasn(\zeta_{a})}\right) \otimes M\left(\sqrt{\jae(\zeta_{b}) \jasn(\zeta_{b}})\right)
    \right]; \quad
 M(\omega)=\left(\begin{matrix}
\omega & 0 \\
0 & 1 
\end{matrix}\right) ,
\end{align}
where $\jae(u)= \jacn(u;k)+\mi \jasn(u;k)$ is defined in terms of the Jacobi elliptic functions: $\jasn(u)=\jasn(u;k),\, \jacn(u)=\jacn(u;k)$, and $k$ is the elliptic modulus.
The resulting $R^{(g)}$ is not symmetric, and has the following form:
\begin{align}
\left(R^{(g)}\right)^{11}_{11}&=a(u_{ab},\zeta_{a},\zeta_{b};k)=\rho [1-\jae(u_{ab})\jae(\zeta_{a})\jae(\zeta_{b})],\notag\\
\left(R^{(g)}\right)^{22}_{22}&=a^{\prime}(u_{ab},\zeta_{a},\zeta_{b};k)=\rho [\jae(u_{ab})-\jae(\zeta_{a})\jae(\zeta_{b})],\notag\\
\left(R^{(g)}\right)^{12}_{12}&=b(u_{ab},\zeta_{a},\zeta_{b};k)=\rho [\jae(\zeta_{a})-\jae(u_{ab})\jae(\zeta_{b})],\notag\\
\left(R^{(g)}\right)^{21}_{21}&=b^{\prime}(u_{ab},\zeta_{a},\zeta_{b};k)=\rho [\jae(\zeta_{b})-\jae(u_{ab})\jae(\zeta_{a})], \label{app:nots:Rg_ij}\\
\left(R^{(g)}\right)^{21}_{12}&=c(u_{ab},\zeta_{a},\zeta_{b};k)=\rho\left[ 1-\jae(u_{ab})\right]\jasn^{-1}(\nicefrac{u_{ab}}{2})\jae(\zeta_{a})\jasn(\zeta_{a}),\notag\\
\left(R^{(g)}\right)^{12}_{21}&=c^{\prime}(u_{ab},\zeta_{a},\zeta_{b};k)=\rho\left[ 1-\jae(u_{ab})\right]\jasn^{-1}(\nicefrac{u_{ab}}{2})\jae(\zeta_{b})\jasn(\zeta_{b}),\notag\\
\left(R^{(g)}\right)^{11}_{22}&=d(u_{ab},\zeta_{a},\zeta_{b};k)= - \mi k \rho [\jasn(\nicefrac{u_{ab}}{2})][1+\jae(u_{ab})]\jae(\zeta_{a})\jasn(\zeta_{a})\jae(\zeta_{b})\jasn(\zeta_{b}),\notag\\
\left(R^{(g)}\right)^{22}_{11}&=d^{\prime}(u_{a}-u_{b},\zeta_{a},\zeta_{b};k)=- \mi k \rho [\jasn(\nicefrac{u_{ab}}{2})][1+\jae(u_{ab})],\notag
\end{align}
where $\rho=\rho(u_{ab},\zeta_{a},\zeta_{b})$ is an arbitrary function. 
The functions in the fermionic $R$-operator \eqref{intro:Rop_orig} are related to the above coefficients $\left(R^{(g)}\right)^{ij}_{km}$ as follows:
\begin{align}
    a_{0}(u_{ab},\zeta_{a},\zeta_{b};k)&=a^{\prime}(u_{ab},\zeta_{a},\zeta_{b};k),\notag \\
    a_{1}(u_{ab},\zeta_{a},\zeta_{b};k)&=b(u_{ab},\zeta_{a},\zeta_{b};k) - a^{\prime}(u_{ab},\zeta_{a},\zeta_{b};k),\notag \\
    a_{2}(u_{ab},\zeta_{a},\zeta_{b};k)&=b^{\prime}(u_{ab},\zeta_{a},\zeta_{b};k) - a^{\prime}(u_{ab},\zeta_{a},\zeta_{b};k),\notag \\
    a_{3}(u_{ab},\zeta_{a},\zeta_{b};k)&=a^{\prime}(u_{ab},\zeta_{a},\zeta_{b};k) - a(u_{ab},\zeta_{a},\zeta_{b};k)-  b(u_{ab},\zeta_{a},\zeta_{b};k) - b^{\prime}(u_{ab},\zeta_{a},\zeta_{b};k),\label{app:nots:a0_d2} \\
    c_{1}(u_{ab},\zeta_{a},\zeta_{b};k)&=c(u_{ab},\zeta_{a},\zeta_{b};k),\notag \\
    c_{2}(u_{ab},\zeta_{a},\zeta_{b};k)&=c^{\prime}(u_{ab},\zeta_{a},\zeta_{b};k),\notag \\
    d_{1}(u_{ab},\zeta_{a},\zeta_{b};k)&=-d(u_{ab},\zeta_{a},\zeta_{b};k),\notag \\
    d_{2}(u_{ab},\zeta_{a},\zeta_{b};k)&=-d^{\prime}(u_{ab},\zeta_{a},\zeta_{b};k).\notag
\end{align}

\newpage
\section{List of equations for $c_{ab}$}
\label{app:eqs}
\setcounter{equation}{0}
\renewcommand\theequation{B.\arabic{equation}}
Here we give the set of independent equations for $c_{ab}$ coefficients in \eqref{ext:two_spin_extension}, that follow from the Yang-Baxter equation \eqref{ext:YBE_two_spin} for the case $u_{a}=u_{b}=u_{c}\equiv u$. We use below the following abbreviations: $\ccos[x]=\cos[x]$, $\ssin[x]=\sin[x]$:

\begin{align}
    &\left\{-c_{ac}+c_{ab}c_{ac}c_{bc}\right\}\ccos\left[\zeta_{a}-\frac{\zeta_{b}+3\zeta_{c}}{2}\right]+\left\{2c_{ac}+c_{bc}-c_{ab}c_{ac}c_{bc}\right\}\ccos\left[\frac{\zeta_{b}-3\zeta_{c}}{2}\right]+\left\{c_{ac}+c_{bc}\right\}\ccos\left[\zeta_{a}-\frac{3\zeta_{b}+\zeta_{c}}{2}\right]\notag\\ 
    &+\left\{-c_{bc}+c_{ab}c_{ac}c_{bc}\right\}\ccos\left[\zeta_{a}+\frac{\zeta_{b}-\zeta_{c}}{2}\right]+\left\{2c_{ab}+c_{bc}-c_{ab}c_{ac}c_{bc}\right\}\ccos\left[\frac{3\zeta_{b}-\zeta_{c}}{2}\right]+\left\{c_{ab}+c_{bc}\right\}\ccos\left[\zeta_{a}+\frac{\zeta_{b}+3\zeta_{c}}{2}\right]\notag \\
    &+\left\{c_{ab}+c_{ac}-2c_{ab}c_{ac}c_{bc}\right\}\ccos\left[\zeta_{a}-\frac{\zeta_{b}-\zeta_{c}}{2}\right]+\left\{-2c_{ab}-2c_{ac}-c_{bc}+c_{ab}c_{ac}c_{bc}\right\}\ccos\left[\frac{\zeta_{b}+\zeta_{c}}{2}\right]\notag \\
    &+\left\{-c_{ab}+c_{ab}c_{ac}c_{bc}\right\}\ccos\left[\zeta_{a}+\frac{3\zeta_{b}+\zeta_{c}}{2}\right]+\left\{-c_{ab}-c_{ac}-c_{bc}-c_{ab}c_{ac}c_{bc}\right\}\ccos\left[\zeta_{a}-\frac{3\zeta_{b}-3\zeta_{c}}{2}\right]\notag \\
    &\left\{-c_{bc}+c_{ab}c_{ac}c_{bc}\right\}\ccos\left[\frac{3\zeta_{b}+3\zeta_{c}}{2}\right]=0, \label{app:eqs:eq1}\\ \newline \notag \\
     &\left\{-c_{ac}+c_{ab}c_{ac}c_{bc}\right\}\ccos\left[\frac{3\zeta_{a}-3\zeta_{c}}{2}\right]+\left\{-c_{bc}+c_{ab}c_{ac}c_{bc}\right\}\ccos\left[\zeta_{b}-\frac{\zeta_{a}-3\zeta_{c}}{2}\right]+\left\{c_{ab}+c_{ac}\right\}\ccos\left[\zeta_{b}+\frac{\zeta_{a}-3\zeta_{c}}{2}\right]\notag\\ 
    &+\left\{-2c_{ab}-c_{ac}-2c_{bc}+c_{ab}c_{ac}c_{bc}\right\}\ccos\left[\frac{\zeta_{a}-\zeta_{c}}{2}\right]+ \left\{-2c_{ab}+c_{ac}-c_{ab}c_{ac}c_{bc}\right\}\ccos\left[\frac{3\zeta_{a}+\zeta_{c}}{2}\right]\notag \\
    &+ \left\{-c_{ac}+2c_{bc}-c_{ab}c_{ac}c_{bc}\right\}\ccos\left[\frac{\zeta_{a}+3\zeta_{c}}{2}\right]+\left\{-c_{ab}+c_{bc}-2c_{ab}c_{ac}c_{bc}\right\}\ccos\left[\zeta_{b}-\frac{\zeta_{a}+\zeta_{c}}{2}\right] \notag\\
       &+ \left\{c_{ac}+c_{bc}\right\}\ccos\left[\zeta_{b}-\frac{3\zeta_{a}+\zeta_{c}}{2}\right] + \left\{-c_{ac}+c_{ab}c_{ac}c_{bc}\right\}\ccos\left[\zeta_{b}+\frac{3\zeta_{a}+\zeta_{c}}{2}\right]+\left\{-c_{ab}+c_{ab}c_{ac}c_{bc}\right\}\ccos\left[\zeta_{b}+\frac{3\zeta_{a}+\zeta_{c}}{2}\right]\notag \\
       &+\left\{-c_{ab}-c_{ac}-c_{ab}c_{ac}c_{bc}\right\}\ccos\left[\zeta_{b}-\frac{3\zeta_{a}-3\zeta_{c}}{2}\right]=0, \label{app:eqs:eq2}\\ \newline \notag \\
       &\left\{-c_{ab}+c_{ab}c_{ac}c_{bc}\right\}\ccos\left[\frac{\zeta_{a}+\zeta_{b}}{2}\right]+\left\{c_{ab}+c_{ab}c_{ac}c_{bc}\right\}\ccos\left[\frac{\zeta_{a}-\zeta_{b}}{2}\right]\ccos\left[\zeta_{c}\right]+\left\{-c_{ab}-c_{bc}\right\}\ssin\left[\frac{\zeta_{a}-\zeta_{b}}{2}\right]\ssin\left[\zeta_{c}\right]=0,\label{app:eqs:eq3} \\ \newline \notag \\
       &\left\{c_{ab}+2c_{bc}-c_{ab}c_{ac}c_{bc}\right\}\ccos\left[\frac{\zeta_{a}-3\zeta_{b}}{2}\right]+\left\{-c_{ab}+c_{ab}c_{ac}c_{bc}\right\}\ccos\left[\frac{3\zeta_{a}+3\zeta_{b}}{2}\right]+\left\{c_{ab}+c_{ac}\right\}\ccos\left[\zeta_{c}-\frac{\zeta_{a}-3\zeta_{b}}{2}\right] \notag \\
       &+\left\{-c_{ab}-2c_{ac}-2c_{bc}+c_{ab}c_{ac}c_{bc}\right\}\ccos\left[\frac{\zeta_{a}+\zeta_{b}}{2}\right]+\left\{c_{ab}+2c_{ac}-c_{ab}c_{ac}c_{bc}\right\}\ccos\left[\frac{3\zeta_{a}-\zeta_{b}}{2}\right]\notag \\
       &+\left\{-c_{ab}+c_{ac}c_{ab}c_{ac}c_{bc}\right\}\ccos\left[\zeta_{c}-\frac{\zeta_{a}-\zeta_{b}}{2}\right]+\left\{-c_{ac}+c_{ac}c_{ab}c_{ac}c_{bc}\right\}\ccos\left[\zeta_{c}-\frac{3\zeta_{a}+\zeta_{b}}{2}\right]\notag \\
       &+\left\{-c_{ab}-c_{ac}-c_{bc}-c_{ab}c_{ac}c_{bc}\right\}\ccos\left[\zeta_{c}+\frac{3\zeta_{a}-3\zeta_{b}}{2}\right]+\left\{c_{ac}+c_{bc}-2c_{ab}c_{ac}c_{bc}\right\}\ccos\left[\zeta_{c}+\frac{\zeta_{a}-\zeta_{b}}{2}\right]\notag \\
       &+\left\{c_{ab}+c_{bc}\right\}\ccos\left[\zeta_{c}+\frac{3\zeta_{a}+\zeta_{b}}{2}\right]=0,\label{app:eqs:eq4} \\ \newline \notag \\
       &\left\{-c_{ac}+c_{ab}c_{ac}c_{bc}\right\}\ccos\left[\frac{\zeta_{a}-\zeta_{c}}{2}\right]+\left\{c_{ac}+c_{ab}c_{ac}c_{bc}\right\}\ccos\left[\frac{\zeta_{a}+\zeta_{c}}{2}\right]\ccos\left[\zeta_{b}\right]+\left\{c_{ab}+c_{bc}\right\}\ssin\left[\frac{\zeta_{a}+\zeta_{c}}{2}\right]\ssin\left[\zeta_{b}\right]=0,\label{app:eqs:eq5} \\ \newline \notag \\
       &\left\{-c_{ab}-c_{ac}+c_{bc}+c_{ab}c_{ac}c_{bc}\right\}\ccos\left[\zeta_{a}+\frac{\zeta_{b}-\zeta_{c}}{2}\right]+\left\{c_{ab}+c_{ac}+c_{bc}+c_{ab}c_{ac}c_{bc}\right\}\ccos\left[\zeta_{a}-\frac{\zeta_{b}-\zeta_{c}}{2}\right]\notag \\
       &+\left\{-2c_{bc}+c_{ac}+2c_{ab}c_{ac}c_{bc}\right\}\ccos\left[\frac{\zeta_{b}+\zeta_{c}}{2}\right]=0.\label{app:eqs:eq6}
\end{align}
\bibliographystyle{elsarticle-num}
\bibliography{zam_hub}
\end{document}